\documentclass[9pt]{extarticle}
\usepackage{spconf,amsmath,graphicx}
\usepackage{amssymb}
\usepackage{multirow}
\usepackage{booktabs}
\usepackage{ctable}
\usepackage{stix}
\usepackage{cite}

\title{MossFormer: Pushing the Performance Limit of Monaural Speech Separation using Gated Single-head Transformer with Convolution-augmented Joint Self-Attentions}
\name{Shengkui Zhao, Bin Ma}
\address{Alibaba Group\\
	\{shengkui.zhao, b.ma\}@alibaba-inc.com\\}
\begin{document}
%
\maketitle
\begin{abstract}
Transformer based models have provided significant performance improvements in monaural speech separation. However, there is still a performance gap compared to a recent proposed upper bound. The major limitation of the current dual-path Transformer models is the inefficient modelling of long-range elemental interactions and local feature patterns. In this work, we achieve the upper bound by proposing a gated single-head transformer architecture with convolution-augmented joint self-attentions, named \textit{MossFormer} (\textit{Mo}naural \textit{s}peech \textit{s}eparation Trans\textit{Former}).  To effectively solve the indirect elemental interactions across chunks in the dual-path architecture, MossFormer employs a joint local and global self-attention architecture that simultaneously performs a full-computation self-attention on local chunks and a linearised low-cost self-attention over the full sequence. The joint attention enables MossFormer model full-sequence elemental interaction directly. In addition, we employ a powerful attentive gating mechanism with simplified single-head self-attentions. Besides the attentive long-range modelling, we also augment MossFormer with convolutions for the position-wise local pattern modelling. As a consequence, MossFormer significantly outperforms the previous models and achieves the state-of-the-art results on  WSJ0-2/3mix and WHAM!/WHAMR! benchmarks. Our model achieves the SI-SDRi upper bound of 21.2 dB on WSJ0-3mix and only 0.3 dB below the upper bound of 23.1 dB on WSJ0-2mix.
\end{abstract}   
\section{Introduction}
\label{sec:intro}

Monaural speech separation that aims to separate individual source speeches from a single overlapped mixture is a fundamental and important task. 
Recent end-to-end deep learning speech separation models have seen large performance improvements  \cite{Luo2019N,Luo2020Z,Nachmani2020Y,Chen2020Q,Liu2019D,Tzinis2020Z,Zeghidour2021D,Subakan2021M}. The time-domain  Conv-TasNet  \cite{Luo2019N} modelled on an encoded representation eventually surpasses the time-frequency domain counterparts. DPRNN \cite{Luo2020Z} provides an effective dual-path framework for handling extreme long encoded input sequences by splitting into smaller chunks and processing the intra- and inter-chunk separately. With capability of learning long-term temporal dependency, DPRNN outperforms Conv-TasNet with a big margin. Building on the dual-path architecture, VSUNOS \cite{Nachmani2020Y} proposes gated RNN modules to further improve the separation performance. However, RNN based models inherently pass history information recurrently through many intermediate states, leading to suboptimal performance. 
\begin{figure}[t]
  \centering
  \includegraphics[width=7.2cm]{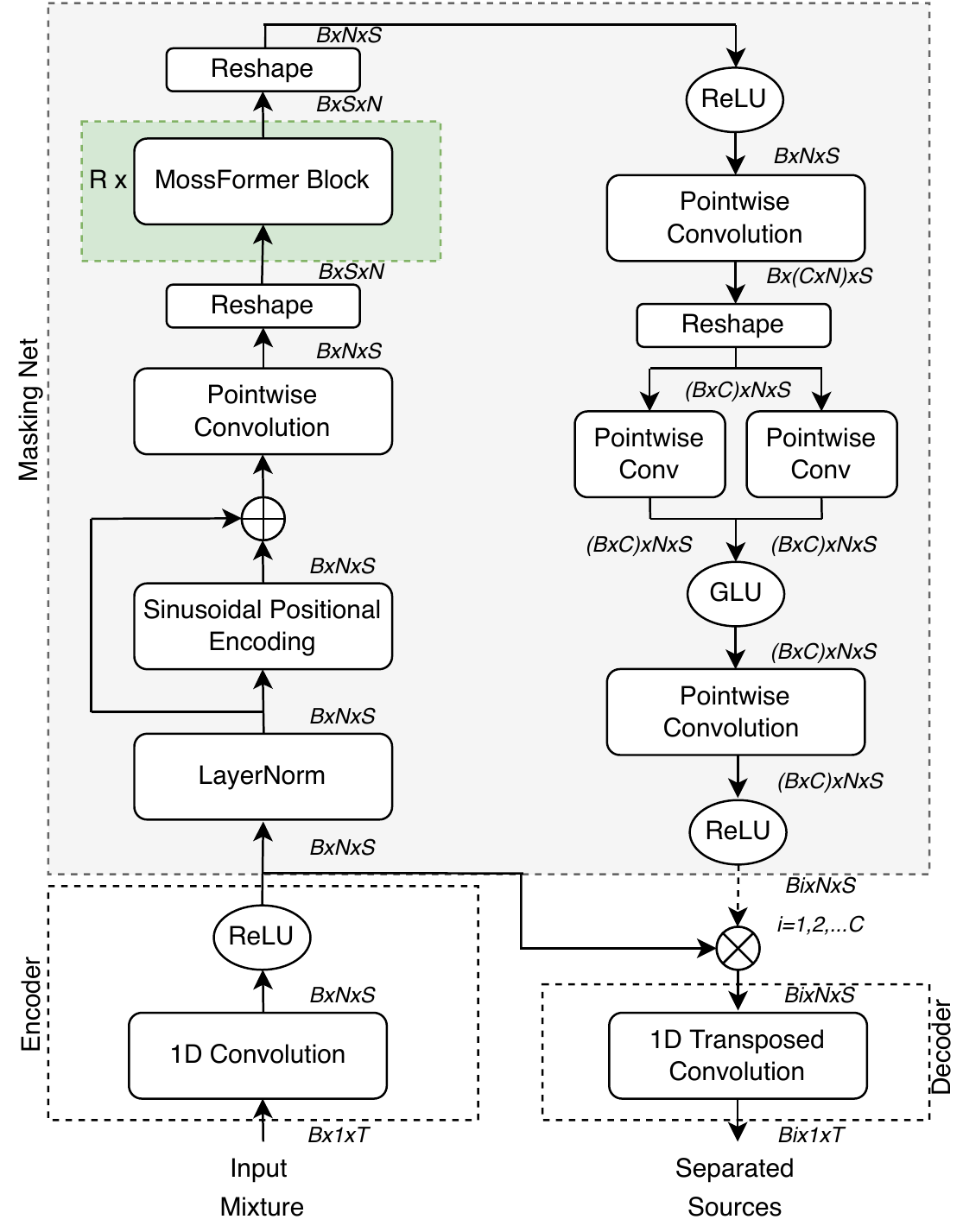}
  \caption{MossFormer model architecture. MossFormer comprises of a convolutional encoder-decoder structure and a masking net. The masking net is a stack of MossFormer blocks and convolutions.}
  \label{fig1}
\end{figure}
Recently, the Transformer architecture based on self-attention \cite{Vaswani2017N} has been successfully integrated into the dual-path speech separation pipeline. Unlike the recurrent learning of RNN, Transformer provide ability to capture long-range elemental interactions directly. DPTNet \cite{Chen2020Q}  uses an amended Transformer architecture with embedded RNN for preserving sequence positional information and shows superior performance than DPRNN. SepFormer \cite{Subakan2021M} completely eliminates RNN recurrence by building on standard Transformer with multi-head self-attention (MHSA) and achieves the state-of-the-art (SOTA) performance. Due to the quadratic complexity over the input sequence in attention computations, the self-attentions in DPTNet and SepFormer are still limited to short context size. The long-range elemental dependencies across chunks are still implicitly modelled through intermediate states. This fact may impose negative impacts on long-range modelling capability. Compared to the recent Cramer Rao bound for non-linear methods \cite{Lutati2022E}, there is still a large performance gap. In addition, convolutions are not well exploited in the existing dual-path Transformer models to learn local feature patterns.  

In this work, we propose a novel \textit{Mo}naural \textit{s}peech \textit{s}eparation Trans\textit{Former} (\textit{MossFormer}) model as illustrated in Figure \ref{fig1}. For the dominant MossFormer block, we propose a gated single-head transformer (GSHT) architecture with convolution-augmented joint self-attentions as illustrated in Figure \ref{fig2}. The GSHT architecture employs a powerful attentive gating mechanism such that only a weakened single-head self-attention (SHSA) is required.  This facilitates a joint local and global self-attention for effective long-range direct interaction modelling. To further model position-wise local feature patterns, we propose a convolution module as illustrated in Figure \ref{fig3} and integrate the convolution module into the attentive gating architecture. Our work is mainly motivated by \cite{gulati20J, Hua2022Z}. 
Our proposed model outperforms SepFormer and the other previous models, and redefines the state-of-the-art on the WSJ0-2/3mix and WHAM!/WHAMR! benchmarks. Moreover, we achieve the SI-SDRi upper bound \cite{Lutati2022E} on WSJ0-3mix. 

\section{The MossFormer Model}
Given a speech mixture $x=\sum_{i=1}^{C}s_i$, we aim to estimate $C$ individual sources $s_i\in\mathbb{R}^{1xT}, i=1,2,\dots,C$ based on a deep learning model. Our overall model architecture is built on the time-domain masking-net framework \cite{Luo2019N} as illustrated in Figure \ref{fig1}. It comprises of a convolutional encoder-decoder structure and a masking net. The encoder-decoder structure responds for feature extraction and waveform reconstruction. The masking net maps the encoded output to a group of masks. 
\subsection{Encoder and Decoder}
The encoder responds for feature extraction and consists of a one-dimensional (1D) convolutional layer (Conv1D) and a rectified linear unit (ReLU), which constrains the encoded output to be non-negative values. 
Let the kernel size of the encoder be $K_1$ with stride of $K_1/2$ and the number of filters be $N$. The input sequence $\mathbf{X}\in\mathbb{R}^{B\times1\times T}$ is encoded to the output $\mathbf{X}^\prime$ as follows:
\begin{equation}
\mathbf{X}^\prime = \mathrm{ReLU}(\mathrm{Conv1D}(\mathbf{X}))
\end{equation}     
where $\mathbf{X}^\prime\in\mathbb{R}^{B\times N\times S}$ and $S=2(T-K_1)/K_1+1$. The batch size $B$ is omitted in the follows for ease of presentation. The sequence $\mathbf{X}^\prime$ is multiplied element-wisely by each individual speaker's mask to obtain the separated feature sequence: $\mathbf{X}_{i}^\dprime=\mathbf{M}_i\otimes \mathbf{X}^\prime$. The separated feature sequence is finally decoded into waveform by the decoder:
\begin{equation}
\hat{s}_i = \mathrm{Transposed\_Conv1D}(\mathbf{X}_{i}^\dprime)
\end{equation} 
The decoder is a 1D transposed convolutional layer and it uses the same  kernel size and stride as the encoder. 
\begin{figure}[t]
  \centering
  \includegraphics[width=5.5cm]{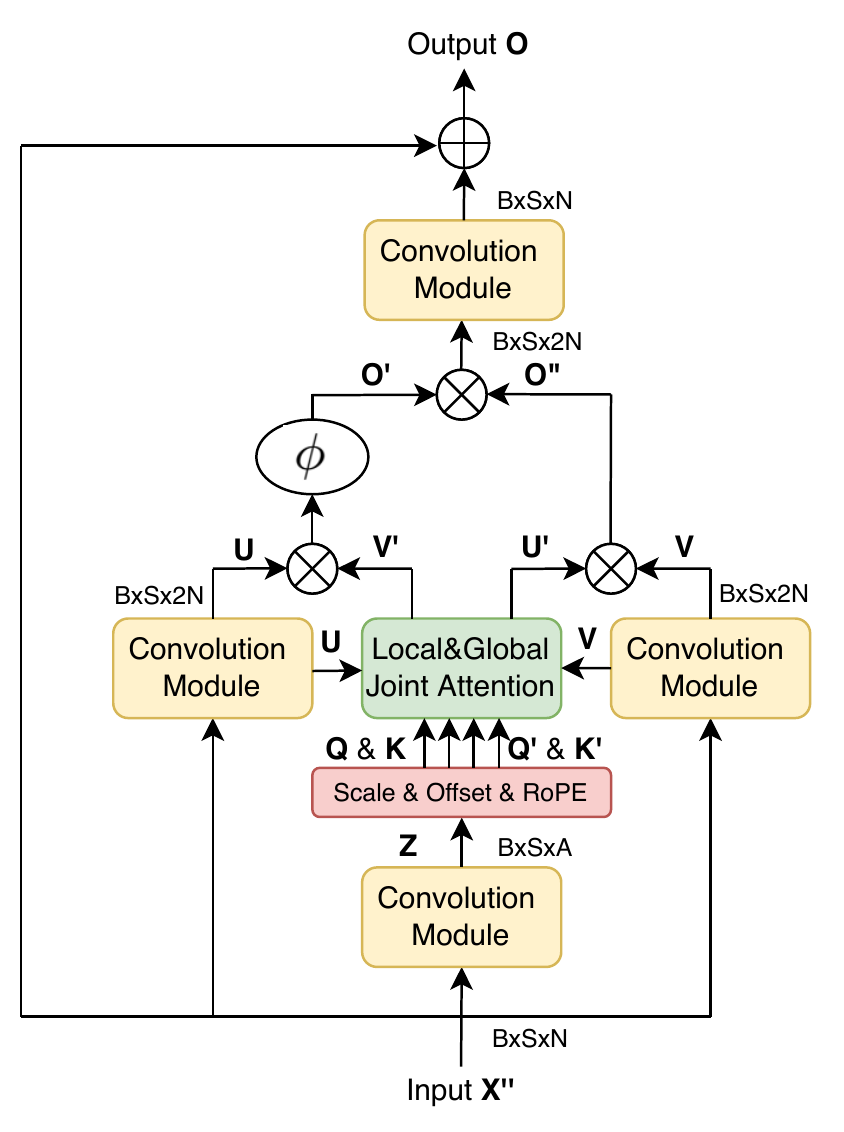}
  \caption{MossFormer block. It consists of four convolution modules, scale and offset operations, a joint local and global single-head self-attention, and three element-wise gating operations ($\phi$ is an element-wise activation function, $\oplus$ is element-wise summation, $\otimes$  is element-wise multiplication).}
  \label{fig2}
\end{figure} 
\subsection{Masking Net}   
The masking net performs a non-linear mapping from the encoder output to $C$ groups of masks as shown in Figure \ref{fig1}. To achieve this, the encoded sequence $\mathbf{X}^\prime$ is first normalized and added with positional encodings for global order information. And the sequence is then passed through a pointwise convolution and after reshaping passed to the MossFormer block for sequential processing. 

In the MossFormer block, the sequence is processed by the convolution modules and the attentive gating mechanism. The  convolution modules process the sequence with linear projections and depthwise convolutions. The attentive gating mechanism performs a joint local and global self-attention and gating operations. 
The MossFormer block learns only the residual and applies the skip connection from the input for ease of training. The output of the current MossFormer block is fed as input to the next MossFormer block. The process of the MossFormer block is repeated $R$ times.  

The output of the final MossFormer block is processed by a ReLU followed by another pointwise convolution, which expands the sequence dimension from $\mathbb{R}^{N\times S}$ to $\mathbb{R}^{(C\times N)\times S}$. It is then passed through a parallel pointwise convolutions and a GLU. Finally, the sequence is passed through pointwise convolution one more time followed by a ReLU to obtain the mask sequence $\mathbf{M}\in \mathbb{R}^{C\times N \times S}$. The mask sequence $\mathbf{M}$ is reformed for each individual speaker $\mathbf{M}_i\in \mathbb{R}^{N\times S}$  and is then fed to the decoder separately. 

\subsection{MossFormer Block} 
The architecture of our proposed MossFormer block is shown in Figure \ref{fig2}, which is developed based on the recent proposed gated attention unit (GAU) \cite{Hua2022Z} for long sequence modelling. A MossFormer block comprises of four convolution modules, scale and offset operations, a joint local and global SHSA, and three gating operations. We aim to boost the modelling capability of the MossFormer block by incorporating convolution modules and a triple-gating structure. The use of gates allows a much simpler SHSA that facilitates a joint local and global attention for effective long-range modelling. 
\subsubsection{Convolution Module}
\begin{figure*}[t]
  \centering
  \includegraphics[width=15cm]{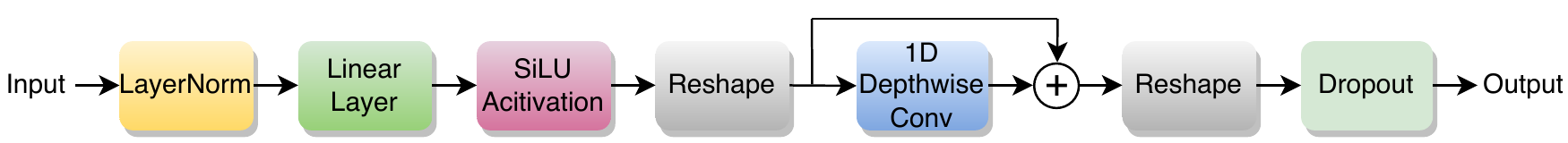}
  \caption{Convolution module. It contains a linear layer with an expansion factor followed by a SiLU activation layer, and then followed by a 1-D depthwise convolution layer with a skip connection.}
  \label{fig3}
\end{figure*}

We propose a convolution module to replace the dense layers in GAU for extracting fine-grained local feature patterns in the MossFormer block. Figure \ref{fig3} illustrates the architecture of the convolution module. In the convolution module, the sequence is first normalized and projected by a linear layer followed by a SiLU. And then it is feature-wise convoluted by 1D depthwise convolution.  Skip-connection and dropout are used to help training and regularizing the network. 
\subsubsection{Attentive Gating Mechanism} 
Our attentive gating mechanism combines attention into the triple-gating process to enhance model capability. The mechanism is formulated as follows. Let the input sequence of the current MossFormer block be $\mathbf{X}^\dprime\in \mathbb{R}^{S\times N}$. It is processed by the convolution module to obtain the values $\mathbf{U}\in \mathbb{R}^{S\times 2N}$ and $\mathbf{V}\in \mathbb{R}^{S\times 2N}$ as follows:
\begin{equation}
\mathbf{U} = \mathrm{ConvM}(\mathbf{X}^\dprime)\mathrm{,}\quad \mathbf{V} = \mathrm{ConvM}(\mathbf{X}^\dprime)
\end{equation}
where ConvM refers to the convolution module. Here we increase the feature dimensions from $N$ to $2N$ in the linear layer of ConvM using an expansion factor of 2. By denoting the attention matrix as $\mathbf{A}\in\mathbb{R}^{S\times S}$, the output sequence $\mathbf{O}\in\mathbb{R}^{S\times N}$ of the MossFormer block can be expressed as follows:
\begin{align}
&\mathbf{O}^\prime=\phi(\mathbf{U}\otimes \mathbf{V}^\prime) \quad\mathrm{where}\quad \mathbf{V}^\prime=\mathbf{AV} \\
&\mathbf{O}^\dprime=\mathbf{U}^\prime\otimes \mathbf{V} \quad\mathrm{where} \quad\mathbf{U}^\prime=\mathbf{AU} \\
&\mathbf{O}=\mathbf{X}^\dprime + \mathrm{ConvM}(\mathbf{O}^\prime\otimes \mathbf{O}^\dprime) 
\end{align}
where the linear layer of ConvM decreases the feature dimension from $2N$ to $N$, and $\phi$ is an element-wise activation function. 
\subsubsection{Joint Local and Global Single-Head Self-Attention} 
For long sequence where $S$ is large, computing the attentions in Eqs. (4) and (5)
directly is very expensive. Fortunately, the presence of gating allows us 
compute $\mathbf{V}^\prime$ and $\mathbf{U}^\prime$ based on the joint local and global attention in an efficient and effective way. 
We first project the input sequence $\mathbf{X}^\dprime$ through the convolution module into a shared representation: $\mathbf{Z}=\mathrm{ConvM}(\mathbf{X}^\dprime)\in\mathbb{R}^{S\times D}$, where $D\ll N$. We then apply low-cost per-dim scalars and offsets, and  RoPE \cite{Su2021Y} to the shared $\mathbf{Z}$ to obtain the queries $\mathbf{Q}$, $\mathbf{Q}^\prime\in\mathbb{R}^{S\times D}$ and the keys $\mathbf{K}$, $\mathbf{K}^\prime\in\mathbb{R}^{S\times D}$ for both the local  and  global attentions. For the global attention, we employ the following low-cost linearised form to capture long-range global interactions for both sequences $\mathbf{V}$ and $\mathbf{U}$:
\begin{equation}
\mathbf{V}_{\mathrm{global}}^\prime=\mathbf{Q}^\prime\left(\beta{\mathbf{K}^\prime}^{T}\mathbf{V}\right)\mathrm{,}\quad \mathbf{U}_{\mathrm{global}}^\prime=\mathbf{Q}^\prime\left({\beta\mathbf{K}^\prime}^{T}\mathbf{U}\right)
\end{equation}
where $\beta=1/S$ is a scaling factor. To compute the local quadratic attention, we chunk $\mathbf{V}$, $\mathbf{U}$, $\mathbf{Q}$, and $\mathbf{K}$ into $H$ non-overlapping chunks of size $P$, where zero-padding is used when $S<H\times P$. The quadratic attention is therefore independently applied to each chunk as follows:
\begin{align}
\mathbf{V}_{\mathrm{local,}h}^\prime=\mathrm{ReLU}^2\left(\gamma\mathbf{Q}_h\mathbf{K}_h^T\right)\mathbf{V}_h\mathrm{,}\enspace
\mathbf{U}_{\mathrm{local,}h}^\prime=\mathrm{ReLU}^2\left(\gamma\mathbf{Q}_h\mathbf{K}_h^T\right)\mathbf{U}_h
\end{align} 
where $\gamma=1/P$ is a scaling factor. Here we adopt the squared ReLU instead of the softmax in MHSA for optimizing performance \cite{Hua2022Z}. Note that $\mathbf{Q}_h\mathbf{K}_h^T$ only needs to be computed once as it is shared by $\mathbf{V}_{\mathrm{local,}h}^\prime$ and $\mathbf{U}_{\mathrm{local,}h}^\prime$. We concatenate all outputs of $\mathbf{V}_{\mathrm{local,}h}^\prime$ and $\mathbf{U}_{\mathrm{local,}h}^\prime$ along the time dimension to form back the full sequences: $\mathbf{V}_{\mathrm{local}}^\prime=[\mathbf{V}_{\mathrm{local,}1}^\prime,\dots, \mathbf{V}_{\mathrm{local,}H}^\prime]$ and $\mathbf{U}_{\mathrm{local}}^\prime=[\mathbf{U}_{\mathrm{local,}1}^\prime,\dots, \mathbf{U}_{\mathrm{local,}H}^\prime]$.  

We add the local attention and the global attention together to form the final joint attentions of $\mathbf{V}^\prime$ and $\mathbf{U}^\prime$ in Eqs. (4) and (5):
\begin{equation}
\mathbf{V}^\prime=\mathbf{V}_{\mathrm{local}}^\prime+\mathbf{V}_{\mathrm{global}}^\prime\mathrm{,}\quad \mathbf{U}^\prime=\mathbf{U}_{\mathrm{local}}^\prime+\mathbf{U}_{\mathrm{global}}^\prime
\end{equation}

\section{Experiments}
\subsection{Dataset}
We evaluate the proposed model on both clean and noisy/reverberated settings using the speech separation benchmarks of WSJ0-2/3mix \cite{Hershey2016Z} and WHAM!/WHAMR! \cite{Wichern2019J, Maciejewski2019G} datasets. We rely on the 8kHz version of the data. The utterances are randomly segmented into 4s long during training and validation. Beside the standard versions of the data, we also consider dynamic mixing (DM) with speed perturbation for data augmentation as described in \cite{Subakan2021M}.  
\begin{table}
\center
\footnotesize
\caption{Hyper-parameters for MossFormer S, M, and L models for optimized performance within parameter and GPU resource limits.}
\begin{tabular}{lccc}
\specialrule{.1em}{.05em}{.05em}
\multirow{2}{*}{Model} &\multicolumn{1}{c}{MossFormer} &\multicolumn{1}{c}{MossFormer}&\multicolumn{1}{c}{MossFormer}        \\
                  & (S)  &(M)       & (L)        \\ \hline 
No. Parameters                    &10.8M & 25.3M  & 42.1M             \\ 
No. MossFormer Blocks ($R$)       &22    & 25     & 24           \\ 
Encoder Output Dimension ($N$)    &256   & 384    & 512               \\ 
Encoder Kernel Size ($K_\mathrm{1}$) / Stride  &8/4   & 16/8   & 16/8   \\
Depthwise Conv Kernel Size ($K_\mathrm{2}$)           &31    & 17     & 17        \\ 
Chunk Size ($P$)                  &256   &256     & 256       \\
Attention Dimension ($D$)   &128   & 128    & 128  \\ 
Gating Activation Function ($\phi$)   &Sigmoid   &Sigmoid     & Sigmoid       \\ \hline 
\specialrule{.1em}{.05em}{.05em}
\label{tab1}
\end{tabular}
\end{table}
\subsection{Training Setup}
Our evaluations are made for three models with parameters of 11M (S), 25M (M), and 42M (L), respectively. The models are chosen based on our best settings of network depth, model dimensions, convolution kernel sizes, chunk size, attention dimensions, as well as gating activation function $\phi$ within parameter size and training resource constraints. We use a single NVIDIA V100 GPU with 16 GB of memory for training. Table \ref{tab1} provides the hyper-parameters.

Our models are implemented based on the SpeechBrain tookit\footnote{speechbrain.github.io/} and  optimized using the SI-SDR training loss \cite{Roux2019S}. We train our models for a maximum of 200 epochs with the Adam optimizer \cite{Kingma2014J} using learning rate of $15e^{-5}$ and batch size of 1. 
During training, the learning rate holds for 85 epochs and then is reduced by a factor of 0.5 with patience of 2. We limit the $l_2$ norm of the training gradients to 5 with gradient clipping. The dropout rate is set to 0.1 for all models. We release audio samples online\footnote{https://github.com/alibabasglab/MossFormer} and source code later.
\begin{table}
\center
\footnotesize
\caption{Performance comparisons on the WSJ0-2mix/3mix and WHAM!/WHAMR! benchmark datasets.}
\begin{tabular}{lccc}
\specialrule{.1em}{.05em}{.05em}
\multirow{2}{*}{Model}  &\multirow{2}{*}{Para.(M)} & \multicolumn{2}{c}{SI-SDRi}         \\
\cline{3-4}
                                         &        & WSJ0-2mix/3mix   & WHAM!/WHAMR!        \\ \hline 
TasNet \cite{Luo2018N}               & -      &   10.8 \enspace\  / \quad -\enspace         & \quad -\quad\   / \quad -\enspace      \\ 
Chimera++ \cite{Wang2018J}         	  & -      &   11.5 \enspace\  / \quad -\enspace         & \ 9.9 \enspace\  / \quad -\enspace            \\
SignPredictionNet \cite{Wang2019K}   & 55.2   &   15.3 \enspace\  / \quad -\enspace         & \quad -\quad\   / \quad -\enspace       \\ 
Conv-TasNet \cite{Luo2019N}          & 5.1    & \enspace 15.3 \enspace\  / \enspace 12.7    & \enspace 12.7 \enspace / \enspace 8.3\enspace              \\ 
DeepCASA \cite{Liu2019D}             & 12.8   &   17.7 \enspace\  / \quad -\enspace         & \quad -\quad\   / \quad -\enspace       \\
Learnable fbank \cite{Pariente2020S} & -      & \quad -\quad\   / \quad -\enspace           & 12.9 \enspace / \quad -\enspace          \\ 
Two-Step CTN \cite{Tzinis2020S}      & 8.6    &   16.1 \enspace\  / \quad -\enspace         & \quad -\quad\   / \quad -\enspace       \\
MGST \cite{Zhao2020C}                & -      &   17.0 \enspace\  / \quad -\enspace         & 13.1 \enspace / \quad -\enspace            \\
DPRNN \cite{Luo2020Z}                & 2.6    & \enspace 18.8 \enspace\  / \enspace 14.7    & \enspace 13.9 \enspace / \enspace 10.3             \\
SuDoRMRF \cite{Tzinis2020Z}          & 2.6    &   18.9 \enspace\  / \quad -\enspace         & \quad -\quad\   / \quad -\enspace       \\ 
VSUNOS \cite{Nachmani2020Y}          & 7.5    & \enspace 20.1 \enspace\  / \enspace 16.9    & \enspace 15.2 \enspace / \enspace 12.2               \\ 
DPTNet \cite{Chen2020Q}              & 2.6    &   20.2 \enspace\  / \quad -\enspace         & \quad -\quad\   / \quad -\enspace       \\ \hline
Wavesplit \cite{Zeghidour2021D}      & 29     & \enspace 21.0 \enspace\  / \enspace 17.3    & \quad -\quad\   / \quad -\enspace         \\ 
Wavesplit + DM                       & 29     & \enspace 22.2 \enspace\  / \enspace 17.8    & \enspace 16.0 \enspace / \enspace 13.2                \\ \hline
SepFormer \cite{Subakan2021M}        & 25.7   & \enspace 20.4 \enspace\  / \enspace 17.6    & \quad -\quad\   / \quad -\enspace       \\ 
SepFormer + DM                       & 25.7   & \enspace 22.3 \enspace\  / \enspace 19.5    & \enspace 16.4 \enspace / \enspace 14.0              \\ \hline 
\textbf{MossFormer(S)}               & 10.8   & \enspace 20.9 \enspace\  / \enspace 17.8    & \quad -\quad\   / \quad -\enspace       \\ 
\textbf{MossFormer(M) + DM}          & 25.3   & \enspace 22.5 \enspace\  / \enspace 20.8    & \enspace 17.1 \enspace / \enspace 15.9                 \\ 
\textbf{MossFormer(L)  + DM}         & 42.1   & \enspace 22.8 \enspace\  / \enspace 21.2    & \enspace 17.3 \enspace / \enspace 16.3     \\ \hline 
Upper bound \cite{Lutati2022E}       & -      & \enspace 23.1 \enspace\  / \enspace 21.2    & \quad -\quad\   / \quad -\enspace       \\ \hline 
\specialrule{.1em}{.05em}{.05em}
\label{tab2}
\end{tabular}
\end{table}
\subsection{Results}
We use SI-SDR improvement (SI-SDRi) as evaluation metric. Table \ref{tab2} reports the results on the clean settings of WSJ0-2mix/3mix datasets and the noisy/reverberated WHAM!/WHAMR! datasets. Our models are compared with the best reported results in the literature. We report results of the small MossFormer(S) model on standard data version and results of the MossFormer(M) and MossFormer(L) models on the augmented data version. For the clean settings, our MossFormer(S) outperforms all the previous models except a very competitive result against Wavesplit on WSJ0-2mix. Note that Wavesplit has 29M parameters, more than 2 times of MossFormer(S), and uses additional speaker identity labels for training. With DM, our MossFormer(M) outperforms all the previous models. On WSJ0-2mix/3mix, MossFormer(M) achieves 22.5 dB and 20.8 dB SI-SNRi compared to 22.3 dB and 19.5 dB for the 26M SepFormer, and 22.2 dB and 17.8 dB for Wavesplit. Our MossFormer(L) makes further performance improvements on top of MossFormer(M) with an increase of the model dimension. Compared to the upper bounds of 23.1 dB and 21.2 dB on WSJ0-2mix/3mix reported in \cite{Lutati2022E}, MossFormer(L) achieves 22.8 dB and 21.2 dB on WSJ0-2mix/3mix. Therefore, MossFormer(L) achieves not only an upper bound  but also new state-of-the-art results on WSJ0-2mix/3mix. 

For the noisy and reverberated settings,  Table \ref{tab2} shows that MossFormer(M) and MossFormer(L) outperform the previous models with big margins and MossFormer(L) achieves new state-of-the-art results on WHAM! and WHAMR!, respectively. For instance, MossFormer(L) achieves 0.9 dB and 2.3 dB more compared to SepFormer. Note that the WHAM!/WHAMR! datasets are built on top of WSJ0-2mix by introducing additional noise and reverberation. Therefore the WHAM!/WHAMR! tasks become harder as the models need to address not only speech separation but also  denoising and dereverberation. We observe that the reverberation affects Wavesplit and SepFormer more than MossFormer as their performance drop more from WHAM! to WHAMR!. 
\subsection{Ablation Studies}
We base MossFormer(S) model and WSJ0-2mix to make ablation studies. Table \ref{tab4} shows the effects of the convolution module, the gating mechanism, and the joint-attention. We observe that the convolution modules has an important impact on the performance. When replacing the convolution modules with the dense layers used in GAU \cite{Hua2022Z} in the values as well as in the queries and keys, it is seen that the performance becomes worse. It also shows that the convolution module affects the queries and keys more than the values. When removing the output $\mathbf{O}^\prime$ in Eq.(6), the triple-gating structure becomes a single-gating structure as in GAU \cite{Hua2022Z} and the result drops from 20.9 dB to 20.4 dB. It shows the effectiveness of the triple-gating design. We also test the model using the quadratic local attention only or the linear global attention only. The results show that none of them performs well individually. It demonstrates the impact of the joint-attention scheme.  Another observation is that the global attention alone performs better than the local attention alone, implying the importance of global modelling.
\begin{table}
\center
\footnotesize
\caption{Ablation studies on the convolution module, the triple-gating mechanism, and the joint attention.}
\begin{tabular}{lc}
\specialrule{.1em}{.05em}{.05em}

Model                           & SI-SDRi        \\ \hline 
MossFormer(S)                    &20.9             \\ 
Quadratic Local Attention Only   &17.8              \\ 
Linear Global Attention Only    &19.6                  \\ 
Remove output $\mathbf{O^{\prime}}$ in Eq.(6)           &20.4          \\
Replace ConvM with Dense for $\mathbf{U}$ \& $\mathbf{V}$  &20.5    \\
Replace ConvM with Dense for $\mathbf{Q}s$ \& $\mathbf{K}s$  &20.3    \\
Replace ConvM with Dense for both $\mathbf{U}$ \& $\mathbf{V}$  and $\mathbf{Q}s$ \& $\mathbf{K}s$ &19.9 \\ \hline 
\specialrule{.1em}{.05em}{.05em}
\label{tab4}
\end{tabular}
\end{table}
\begin{table}[t]
\center
\footnotesize
\caption{Ablation study on different choice of the gating activation function $\phi$. }
\begin{tabular}{lccccc}
\specialrule{.1em}{.05em}{.05em}
\multirow{2}{*}{Model} & \multicolumn{5}{c}{Gating Activation Function $\phi$}         \\
\cline{2-6}
                          &ReLU  & GELU    &Swish   &Bilinear  &Sigmoid  \\ \hline
MossFormer(S)             &20.0   &19.9    &19.9   &20.1      &20.9   \\ 
\hline
\specialrule{.1em}{.05em}{.05em}
\label{tab5}
\end{tabular}
\end{table}

\begin{table}[h]
\center
\footnotesize
\caption{Ablation study on settings of attention dimension and chunk size.}
\begin{tabular}{lccc|ccc|ccc}
\specialrule{.1em}{.05em}{.05em}
\multirow{2}{*}{} & \multicolumn{3}{c}{Kernel Size $K_\mathrm{2}$} & \multicolumn{3}{c}{Attention Dim. $D$}  & \multicolumn{3}{c}{Chunk Size $P$}       \\
\cline{1-4}\cline{5-7}\cline{8-10}
&21     &31    &65     &64      &128      &256    &128    &256   &384  \\ 
&20.6   &20.9  &20.7   &20.5    &20.9     &20.8   &20.5   &20.9  &20.9  \\ 
\hline
\specialrule{.1em}{.05em}{.05em}
\label{tab7}
\end{tabular}
\end{table}  
In Table \ref{tab5}, we validate the choice of the activation function $\phi$ according to the study \cite{Shazeer2020} and it shows that the Sigmoid function works the best. Table \ref{tab7} studies the effects of kernel size $K_2$ in the depthwise convolution in ConvM, the attention dimension $D$, and the chunk size $P$. We find that the performance improves with larger kernel sizes till 31 but worsens for 65. In addition, $D=128$ performs the best. A larger chunk size tends to perform better, but further increasing $P$ from 256 to 384 has no more improvement.  
\section{Conclusions}
In this work, we introduced MossFormer, a Transformer model for monaural speech separation. Contrary to prior models, we learn the local feature patterns and the global long-range dependencies in a unified attentive gating model. Unlike the dual-path framework that models the long-rang interactions implicitly, we employed a quadratic local attention and a low-cost global attention in a joint form to model the long-rang interactions directly. We also combine convolution modules to model the local feature patterns. Our studies demonstrated the importance of each component and achieved much better results than previous models with new state-of-the-art on the benchmarks of WSJ0-2/3mix and WHAM!/WHAMR!. 
\bibliographystyle{IEEEbib}
\bibliography{refs}

\end{document}